\documentclass[aps,twocolumn]{revtex4}
\usepackage[T1]{fontenc}
\usepackage[latin1]{inputenc}
\usepackage{graphicx}
\usepackage{psfrag}
\usepackage{pslatex}



\begin{document}

\begin{widetext}
\begin{quote}
\emph{Draft for:} Proceedings, The Nature of Time: Geometry, Physics \&
Perception;\\
May 21-24, 2002, Tatranska Lomnica, Slovak Republic
\end{quote}
\end{widetext}

\title{Time contortions in modern physics}

\author{A. F. Kracklauer}

\address{\tt{kracklau@fossi.uni-weimar.de}}

\begin{abstract}
As a basis for epistemological study of ``time,'' we analyze three suspect
phenomena introduced by modern physics: non-locality, asymmetric aging and advanced
interaction. It is shown that all three arise in connection with what has to
be taken as arbitrary ideosyncrasies in formulation. It is shown that minor
changes result in internally consistent variations of both Quantum Mechanics
and Special Relativity devoid of these phenomena. The reinterpretation of some
experiments though to confirm the existence of non-locality and asymmetric aging
is briefly considered and a possible test is proposed.

\end{abstract}
\maketitle

\section{Introduction}

Modern Physics, specifically, Quantum Mechanics (QM) and Special Relativity
(SR), have brought three notions into common currency, namely: ``non-locality,''
``asymmetric aging'' and ``advanced interaction,'' that defy accommodation with
classical physics, common sense and, we hold, basic logic. How this came about
historically is well documented in the literature and so will not be belabored
here. Although asymmetric aging, a.k.a. the ``Twin Paradox,'' has been a foil
for `nondomesticated' newcomers and numerous outsiders for 90 years, all of
these notions nowadays enjoy solid ensconcement in the corpus of ``verified
physics.'' One reason for this appears to be their appeal as harbingers from
the preternatural; or as a foot-in-the-door leading out from the stolid, unromantic
material world to a great mystical beyond. Another reason is that too little
effort has been devoted to seeking alternatives out of respect for the profound
successes of both QM and SR. It is this last deficit, we aim to attack.

\section{Instantaneous interaction in QM }

The case \emph{for} instantaneous interaction is based on two arguments. One
was originated by Bell with his now famous derivation of inequalities.\cite{Bel87}
The other exploits arguments pertaining to the algebraic properties of spin
operators. The later argument has a rather more complicated pedigree, but the
key ideas are most often credited to Kochen and Specker.\cite{Mer93}\cite{Koc67}
Bell's main argument, and the drift of all of his analysis, is in the form of
a \emph{reductio ad absurdum.} That is, he argues that by assuming that a fully
local, realistic extention of QM with hidden variables exists, then certain
inequalities derived on the assumption of locality (as he encodes it) constrain
the coincident probabilities for the Einstein, Podolsky and Rosen (EPR) Gedanken
experiment in a way that can be tested. In fact, these inequalities are violated
in experiments, so that one arrives at a empirical contradiction with Bell's
assumptions.\cite{Afr99} 

The case \emph{against} such instantaneous interaction, on the other hand, stands
on three legs. The first consists of detailed analysis of the logical structure
of the basic hypotheses supporting the existence of non-locality. Such arguments
turn mainly on minutia pertaining to the definitions of statistical statements
and their application to EPR Gedanken-type experiments. The second leg goes
straight to constructing counterexamples. If a valid, classical model for EPR
experiments can be found --- in direct conflict with the conclusions of Bell's
analysis --- then obviously the hypotheses going into this analysis must be
wrong even when the exact error remains unidentified. Such models can consist
of either calculations of plausible setups, or ``Monte Carlo'' simulations of
such setups. The final leg consists of experiments in non traditional regimes.
Optical experiments based on EPR considerations have most often been done in
the visible part of the spectrum where ``photon'' phenomena are most evident.
But, because the reasoning behind the Bohm variation of EPR setups is not dependant
on the wave length, these same setups can be executed in other parts of the
EM spectrum in which the peculiarities of ``photodetectors'' and therefore
``photons'' are irrelevant.

\subsection{Fundamental errors}

To the best of this writer's information, the fundamental error in Bell's reasoning
was first identified by Jaynes.\cite{Jay89} (There are, in addition, those
who claim that the point made by Jaynes was used already before Bell's seminal
paper. In other words, had Bell focused on such work in connection with his
analysis of EPR correlations, he never would have authored his now famous paper.)
Subsequent to Jaynes, the kernel of his observation was rediscovered independently
in various styles at least three times.\cite{Per91}-\cite{Hes01} 

The core of Jaynes' point is that Bell misapplied Bayes' formula. The marginal
probability distribution of a coincident probability dependant on three variables
takes the form
\begin{equation}
\label{coincidence}
P(a,\, b)=\int P(a,\, b,\, \lambda )d\lambda \, .
\end{equation}
By basic probability theory, the integrand of this equation is to be decomposed
in terms of individual detections in each arm according to Bayes' formula
\begin{equation}
\label{bell2}
P(a,\, b,\, \lambda )=P(\lambda )P(a|\, \lambda )P(b|a,\, \lambda ),
\end{equation}
where \( P(a|\, \lambda ) \) is a conditional probability. In turn, the integrand
of Eq. (\ref{coincidence}) can be converted for use in integrand of Bell's
\emph{Ansatz} in which he considers the correlation of EPR variables:
\begin{equation}
\label{coinprob}
Corr(a,\, b)=\int aP(a,\, \lambda )\: bP(b,\, \lambda )\rho (\lambda )d\lambda ,
\end{equation}
iff
\begin{equation}
\label{nux}
P(b|\, a,\, \lambda )\equiv P(b|\, \lambda ),\; \, \, \forall a.
\end{equation}
 (In Bell's notation, \( A[B](a[b],\: \lambda )=a[b]P(a[b],\: \lambda ) \) where
 all \(P\)'s are probabilities corresponding to the moduli of wave functions;
in accord with current practice, even better notation would be: \( A(B)(a(b),\: \lambda )=a(b)P(a(b)\: |\lambda ) \).)
Bell argued, in Jaynes' terms, that strict locality implies that the dependence
of \( P(b|\, a,\, \lambda ) \) on \( a \), implies a \emph{causative} relationship
between the measuring stations. This is clearly not the case for Bayes' formula
as the correlation can just as well arise from a \emph{common cause} at a point
in the intersection of the past light cones of both detectors.\cite{Fel50}
Thus, we conclude that Bell's factorization, is just not tenable. 

Bell's analysis would have been standard statistical analysis were is possible
to introduce a \( \lambda  \)-meter. However, if \( \lambda  \) is a `hidden
variable' intrinsically unavailable for observation but whose existence is to
be inferred by observing the patterns in the values revealed by \( a \)- and
\( b \)-meters, then whatever \( \lambda  \) encodes must be evident in those
values, but Bell's encoding of locality precludes this by inadvertent assumption. 

Although this argument against Bell's factorization is clean and indisputable
in the optical case (actually for any EM phenomena), it is unfair to the total
challenge faced by Bell or any physicist with an eye to the consistency of all
of physics. Their problem here is this: particle beams are seen to ``navigate''
as if they had a wave character but register at detectors as if they are composed
of collections of `particles.' That is, the dualistic behavior of particle beams
implies that in some concrete sense that the beams are ontologically ambiguous
while they are underway and resolved only at measurement, in other words, their
wave packet is collapsed. The particle character of waves, on the other hand,
can be attributed to the conversion of the continuous energy stream of the beam
into a digitized photocurrent comprised of electrons. Thus, this ambiguity is
not necessary for radiation beams but seems to have been imposed anyway for
the sake of overall uniformity. The final point of these considerations is,
that without a means of rationalizing beam behavior, for the particle beam equivalent
of coincident probabilities considered by Bell for optical, i.e., radiation
beams, the implied causality relationship must be respected. This is so, because
when one arm of such a wave function is `collapsed' by measurement, then its
partner also collapses at the same instant. In other words, for particle beams,
a wave function should not be just an epistemological \emph{aid d'memoire} but
somehow also substance.

Therefore, Jaynes' argument is incomplete without an accompanying classical
model for particle beam wave-like behavior. In fact, such exists\cite{Kra99};
so his argument stands.

\subsection{Counterexamples, classical models and simulations}

The earliest local realistic model for EPR correlations known to this writer
is Barut's model for the original spin variant of Bohm's rendition of the EPR
setup.\cite{Bar92} Some other attempts were mislead by the presumed correctness
of Bell's arguments and tried simultaneously to satisfy both classical physics
and Bell's inequalities.\cite{San96}\cite{Aer99} These latter models can be
considered rigorous technical counterexamples to Bell's reasoning but they still
fail to be convincing because they also incorporate features that are at odds
with some other seemingly necessary characteristic. A systematic study providing
models for various EPR and higher order (i.e., GHZ) experiments is only recently
published.\cite{Kra02} Herein only the most basic variant is considered as
an illustration.

To model the prototypical EPR experiment with `entangled' polarization states,
the source is assumed to emit a double signal for which individual signal components
are anticorrelated and, because of the fixed orientation of the excitation,
confined to orthogonal polarization modes; i.e.
\begin{equation}
\label{30}
\begin{array}{cc}
S_{1} & =(cos(n\frac{\pi }{2}),\: sin(n\frac{\pi }{2}))\\
S_{2} & =(sin(n\frac{\pi }{2}),\: -cos(n\frac{\pi }{2}))
\end{array},
\end{equation}
 where \( n \) takes on the values \( 0 \) and \( 1 \) with an even, random
distribution. The transition matrix for a polarizer is given by,

\begin{equation}
\label{40}
P(\theta )=\left[ \begin{array}{cc}
\cos ^{2}(\theta ) & \cos (\theta )\sin (\theta )\\
\sin (\theta )\cos (\theta ) & \sin ^{2}(\theta )
\end{array}\right] ,
\end{equation}
 so the fields entering the photodetectors are given by:
\begin{equation}
\label{50}
\begin{array}{cc}
E_{1} & =P(\theta _{1})S_{1}\\
E_{2} & =P(\theta _{2})S_{2}
\end{array}.
\end{equation}
 Coincidence detections among \( N \) photodetectors (here \( N=2 \)) are
proportional to the single time, multiple location second order cross correlation,
i.e.:
\begin{equation}
\label{e2}
P(r_{1},\, r_{2},..r_{N})=\frac{<\prod ^{N}_{n=1}E^{*}(r_{n},\! t)\prod ^{1}_{n=N}E(r_{n},\! t)>}{\prod ^{N}_{n=1}<E^{*}_{n}E_{n}>}.
\end{equation}
 The final result of the above is:
\begin{equation}
\label{60}
P(\theta _{1},\theta _{2})=\frac{1}{2}\sin ^{2}(\theta _{1}-\theta _{2}).
\end{equation}
 This is immediately recognized as the so-called `quantum' result. (Of course,
it is also Malus' Law, thereby being in total accord with one premise of this
report.) 

Likewise, EPR correlations can be simulated fully on the basis of classical
physics.\cite{Hof01}

\subsection{Empirical counterexamples}

The ``quantum'' character of EPR experiments resides in the peculiarities of
``photons.'' With respect to these experiments, however, the view point can
be taken that photons are just a means of accounting for the fact that continuous
radiation is converted in a photodetector to a digitized photocurrent. Thus,
where EPR experiments are done in a part of the spectrum in which it is possible
to track the time evolution of an electric field, then the photocurrent can
be raised to such a high intensity that it can be regarded as a continuous entity
like radiation. The EPR correlations then become simply those among current
intensities. Such experiments have been done and the results are in full conformity
with the so-called ``quantum'' results. Perhaps the first experiment of this
nature was done by Evdokimov et al. with radar gear.\cite{Evi96} Recently,
a four-fold ``GHZ'' experiment using hetrodyning techniques has also been done,
again giving results in full conformity with those from QM and in direct contradiction
with the conclusion of Bell's ``theorem.''\cite{Lee02} In short, these experiments
provide purely classical examples of the origin of EPR correlations.

\section{Asymmetric aging}

In classical mechanics, the 3-D vector position is the dependant variable while
time is an independent parameter. Likewise, in special relativistic mechanics,
the 4-D vector of space-time `location' is the dependant variable and proper
time is the independent parameter, \emph{so long as:} a single particle in a
field is under consideration. When two mutually interacting particles are taken
into consideration, this structure seems to break down because it is held that
the proper time intervals on separate world lines between two crossings are
unequal.\cite{Tru97} The most renown illustration of this situation is known
as: ``the twin paradox.''\cite{Ein05} 

It is, however, the contention herein that this situation is the result of error.
When this error is corrected, asymmetric aging is seen not to occur.\cite{Kra00b}
The cause of the error is found in a nonintuitive property of the Lorentz transformation:
it induces nonuniform scale changes. Although this latter fact is well known,
its effect on what can be called ``space-time'' perspective, is still oft
ill understood and misapplied. 

Customarily analysis of the twin paradox has not carefully taken into account
the determination of the distance to the turn-around point (which for brevity,
we denote the `pylon') of the traveling twin. This distance is not really a
vector on a Minkowski diagram but rather the space-like separation of two entire
world lines, namely those of the terminus and pylon of the trip. The pylon,
that is, its place in the world, is not an event but a location, in other words,
a worldline. The turn-around itself is, of course, an event in the usual meaning
of that word for special relativity. For the traveling twin, however, the turn-around
event \emph{per se} is a secondary matter as far as his navigational needs are
concerned. His primary concern is to travel to the designated point in space,
regardless of the time taken, before reversing course. How can he do this? First,
he and his stay-at-home sibling would chart a course before the beginning of
the trip; that is, they would select an object in the world, a star say, and
designate it as the pylon. From standard references they find that this star
is located in a particular direction at a distance \( D. \) This distance is
not the length of a unique Lorentz vector but the proper length of the displacement
from the home location of the twins; i.e., the length of all space-like Lorentz
vectors connecting these two world lines. For parallel world lines, this value
is invariant starting from any arbitrary point on either world line. With this
in hand, the traveling twin then determines the speed capabilites of his craft
and calculates the anticipated arrival time at the pylon.

The distance to the pylon star is not an apparent distance, the length of a
moving rod as seen from a second frame, for example, but the proper length to
the whole world line of the selected star. Such a length is a scaler and a Lorentz
invariant. The location of the world line of the pylon on a Minkowski diagram
depends on the axis to which it refers. That is, this world line with respect
to the stationary twin passes through the space coordinate at `\( D \)' on
the his abscissa. Likewise, this world line must pass through the traveler's
abscissa also; but, because of the difference in the scale of the traveler's
axis, this same world line, although still parallel to the stay-at-home's world
line, will not be congruent to the line referred to the stay-at-home's axis
but is displaced by the scale factor. (It is this displacement that has been
overlooked in previous analysis and which distinguishes this approach.) The
consequence of this displacement is that, the intersection of the traveler's
world line with the world line of the turn~-around point is also further out
on the traveler's world line; i.e., the proper time taken to reach the pylon
is seen to be greater than heretofore estimated. In fact, it is equal to the
proper time of the stay-at-home as he himself computes it for the time taken
by the traveler to reach the pylon. Thus, when the whole trip is completed,
both twins agree that they have experienced equal portions of proper time since
the start of the trip; i.e., their internal clocks, ages, are equal. Their reports
to each other via light signals on the passage of time, in the usual way do
not agree, however. The final consequence of these considerations is that, contrary
to oft expressed opinion, proper time can serve in a self consistent way as
the independent variable for relativistic mechanics.

These points can be depicted as follows on a Minkowski chart. (See Figure 1)

{\begin{figure*}[ht]
\psfrag{D}[][][0.8]{$D$}   
\psfrag{XZ}[][][0.8]{$x$} 
\psfrag{TZ}[][][0.8]{$t$}
\psfrag{XQ}[][][0.8]{$x'$} 
\psfrag{TQ}[][][0.8]{$t'$} 
\psfrag{XZ, TZ}[][][0.80]{$(x,\,t)$} 
\psfrag{XQ, TQ}[][][0.80]{$(x',\,t')$} 
\includegraphics[width=1.8\columnwidth]{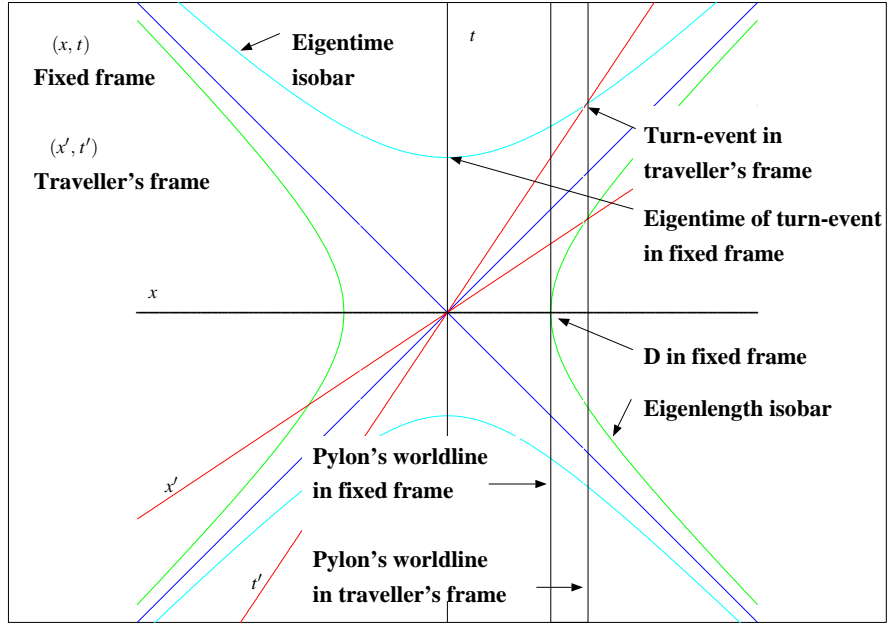}

\caption{This figure is comprised of two Minkowski charts superimposed on top
of each other. The worldline of the Pylon in the fixed frame chart passes
through the point `\( D \)' on the \(x\)-axis. The corresponding point on the
\(x'\)-axis, the traveler's axis, is found by sliding up the proper-length
isocline to the intersection with the \(x'\)-axis. The worldline of the pylon
passes through this point on the prime chart. The intersection of the Pylon's
worldline with the \(t'\)-axis is the point on the traveler's chart
representing the `turn-around' event. The proper-time of the turn-around event
in the fixed frame is found by sliding down that proper-time isocline which
passes through the turn-around event to its intersection with the \(t\)-axis.
It is clear that this value is identical with the time assigned by the fixed
twin himself to the turn-around event as it may be projected horizontally over
to the intersection of the Pylon's worldline in his (fixed) frame with the time
axis of the traveler. Apparent asymmetric ageing arises by using, incorrectly,
that proper-time isocline which passes through the intersection of the
traveler's and the pylon's fixed frame worldlines.}  
\label{Figure}
\end{figure*}}

The same conclusion can be won also as follows: Let \( \mathbf{x}_{j} \) be
the Minkowski configuration four-vector with components \( x_{j},\: y_{j},\: z_{j},\: ict \)
of the \( j \)-th particle. Let \( \mathbf{dx}_{j} \) be a differential displacement
along the \( j \)-th particle's orbit; i.e., a differential of arc length.
Two such differentials tangent to arbitrary points \( p \) and \( p' \) on
orbits \( j \) and \( k \) are related to each other by the Lorentz transformation
\( L(p,\: p',\: j,\: k) \), between the instantaneous rest frames of \( j \)
and \( k \); i.e., given \( \mathbf{dx}_{j}|_{p} \), \( \mathbf{dx}_{j}|_{p'} \)
is essentially defined by
\begin{equation}
\label{2.1}
\mathbf{dx}_{k}|_{p'}=L(p,\: p',\: j,\: k)\mathbf{dx}_{j}|_{p}).
\end{equation}
 Thus, the differential of arc length,\( (\mathbf{dx}_{j}\cdot \mathbf{dx}_{j})^{1/2} \)
is invariant because at each point it satisfies
\begin{eqnarray}
(\mathbf{dx}_{k}|_{p'}\cdot \mathbf{dx}_{k}|_{p'})^{1/2}= & (\mathbf{dx}_{j}|_{p}L^{*}\cdot L\mathbf{dx}_{j}|_{p})^{1/2} & \nonumber \\
= & (\mathbf{dx}_{j}|_{p}\cdot \mathbf{dx}_{j}|_{p})^{1/2}. & \label{2.2} 
\end{eqnarray}
 All such differentials may, therefore be set equal to the common differential
\( c\, d\tau  \), where \( c \) is the speed of light and \( \tau  \) is
the independent parameter which assumes the units of time; i.e., 
\begin{equation}
\label{2.3}
c\, d\tau =(\mathbf{dx}_{j}\cdot \mathbf{dx}_{j})^{1/2}=(\mathbf{dx}_{k}\cdot \mathbf{dx}_{k})^{1/2}.
\end{equation}
 Dividing (2.3) by \( c \) and rewriting yields
\begin{equation}
\label{2.4}
d\tau =dt_{j}\gamma ^{-1}_{j}=dt_{k}\gamma ^{-1}_{k},
\end{equation}
 where \( \gamma ^{-1}_{j}=\left( 1-\left( v_{j}/c\right) ^{2}\right) ^{1/2} \)
in the customary notation. 

Alternately, this conclusion also follows directly from standard formulas. It
is known that all four-velocity vectors satisfy \( \mathbf{v}\cdot \mathbf{v}=c^{2} \),
so that differentiating by \( \tau  \) leads to the conclusion that four-acceleration
is always orthogonal to the four-velocity.\cite{Rin01} This means that acceleration
does not change the modulus of the velocity, in other words, the `four-lengths'
of all velocity vectors for all particles equal each other so that by dividing
each by \( (d\tau )^{2} \) and taking the square root, one obtains Eq. (\ref{2.3})
again.

\subsection{Conflict with experiments}

All standard works on Special Relativity cite experiments attesting to the ``reality''
of time dilation and the consequent aging-discrepancy. How are they to be understood
in view of the above considerations? First, note that to date no experiment
meets the conditions leading to the twin-paradox. Certain experiments, those
involving muon decays, for example, are described by linear transformations
but are not round trips. ``Clocks-around-the-world'' experiments did involve
round trips, but not linear (acceleration free) motion. Further, note  that
time dilation is `real' in the sense that it actually occurs with respect to
signals, and is really no more puzzling than the fact that one's hand (\( \sim 10^{2} \)\( \,  \)cm\( ^{2} \))
can shade the sun (\( \sim 10^{40}\,  \)cm\( ^{2} \)). It is an effect attendant
to `perspective' in space-time. Thus, all physical effects resulting from the
`appearance' (i.e., the way in which light signals transmit information or momentum-energy)
will be modified by the the perspective. So any test of time dilation which
involves a report from, or the interaction between, disparate frames will exhibit
phenomena resulting from relative positions and times of emitter and receiver;
i.e., space-time perspective.

Some experiments seem exempt from the effects of perspective. The two customary
examples are the muon decay curve in the atmosphere, and the transport of atomic
`clocks-around-the-world.' Here the situation is less clear. Each of these experiments,
however, is afflicted with features that allow contest.\cite{Gal97}

Muon decay, for example, largely seems to ignore possible cross-section dependence
on the velocity of the projectile and secondary production. The clocks-around-the-world
experiment has been strongly criticized for its data reduction techniques. In
particular, the existence of time delay effects for transported clocks has been
questioned.\cite{Kel00} Without access to the details of these experiments
and their subsequent data analysis, one is not in position to do deep critical
analysis; nevertheless, there is sufficient information in the literature to
reasonably justify considering conclusions drawn on their basis as disputable.
Moreover, experience with contemporary communication technology seems to present
numerous practical reasons to question the conventional understanding of time
delay effects for transported clocks.\cite{Fla98}

On the other hand, there are also experimental results completely in accord
with this result. An attempt by Phipps to observe the so called Ehrenfest effect---Fitzgerald
contraction of the circumference of a disk as a consequence of high tangential
velocity due to rotation---gave unambiguous null results, for example.\cite{Phi74}
The lack of radial dependence of element abundance and star species in observed
galaxies can be taken as cosmic scale confirmation of Phipps' result.

\subsection{A proposed test}

Crucial to a test of this formulation is that the aging of `twins' be compared
directly rather than via reports conveyed between frames. Because customary
experiments rely on signals sent from the moving to the fixed frame one way
or another, it is not possible to exclude `space-time' perspective effects. 

Perhaps this can be overcome. Consider a variation of a Pound-Rebka experiment
employing a material with an element whose nucleus is naturally unstable. Let
a sample of this material be divided and then hold half at a high temperature
and half at a low temperature long enough such that the calculated time dilation
of the more rapidly moving atoms of the heated half is great enough to yield
a detectable difference in decay products. The ratios of decay products then
should be compared finally in the same frame, i.e., at the same temperature.
An experiment of this structure would not be dependant on the transmission of
signals from frame to frame but simply internally tally the total passage of
eigen time in terms of decay half-lives in each frame for subsequent comparison.
(Note: this scheme can be considered only conceptual inspiration. In fact the
shape of decay curves conceals rather than enhances differences in the accumulation
of proper ages.)

\section{Advanced interaction}

Electrodynamics as field theory (i.e., Maxwell's equations) does not result
in a closed formulation. That is, the interaction between two charged particles
is describe by considering one charge as a current and solving for the fields
at the position of the second which is then `moved' according to the Lorentz
force law. Then, the second particle is considered a source current which generates
perturbing fields back on the first charge. Thereafter the first charge's motion
is corrected and used to recalculate its fields at the position of the second---\emph{ad
infinitum,} or to the desired accuracy.\cite{Roh65}

Fokker developed a closed formulation for the electromagnetic force by incorporating
light-cone into action-at-a-distance mechanics. Essentially he found a Lagrangian
which is not merely the sum of of individual Lagrangians whose variation yields
coupled equations of motion.\cite{Fok29} This Lagrangian, however, produced
yet another complexity: It led to simultaneous advanced and retarded interaction
for each particle. This feature is problematic on two levels. First, it raises
questions of causality because it would mean that the present is always partially
conditioned by all of the future, contrary to observation. Secondly, it introduces
the calculational complication of precluding the known methods of integrating
the equations of motion (this point will be discussed below).

No resolution for the causality difficulties of the \emph{pure} two-particle
problem appear to have been proposed; in fact, apparently the only attempt at
resolution immerses the problem in a many body universe by invoking radiation
absorbers at infinity.\cite{Whe45} Moreover, although integration of the pure
two-particle equations has been attempted, thus far the proposed schemes are
clearly approximation techniques or useful in severely restricted circumstances.

Taking advantage, however, of the integrity of proper-time, \cite{Kra78} we
can formulate direct interaction mechanics as follows: Let four-velocities be
defined as
\begin{equation}
\label{2.6}
\mathbf{v}_{j}:=\mathbf{dx}_{j}/d\tau =\gamma _{j}(\mathbf{v}_{j},\, ic):=\dot{\mathbf{x}}_{j}
\end{equation}
 and momenta as \( m_{j}\mathbf{v}_{j} \), where \( m_{j} \) is the \( j \)-th
particle's rest mass. With these definitions, the four-vector version of Hamilton's
principle
\begin{equation}
\label{2.7}
\delta \int ^{\tau _{2}}_{\tau _{1}}\cal {L}\bf {x_{j},\, v_{j}},\, \tau )d\tau =0,
\end{equation}
 where (for \( N \) (number of particles) =2)
\begin{eqnarray}
{\cal {L}}=\sum ^{2}_{j=1}m_{j}\left( \mathbf{v}_{j}\cdot \mathbf{v}_{j}\right) ^{1/2} &  & \nonumber \\
-2\sum ^{2}_{k\neq j}e_{j}e_{k}\int ^{\tau }_{-\infty }\mathbf{v}_{j}(\tau )\cdot \mathbf{v}_{k}(\tau ')\delta \left( \left( \mathbf{x}_{j}(\tau )-\mathbf{x}_{k}(\tau ')\right) ^{2}\right) d\tau ', &  & \label{2.8} 
\end{eqnarray}
 yields equations of motion coupled by only two interactions (Because of the
upper bound on the integral, advanced interactions are excluded.):

\begin{equation}
\label{2.9}
m_{j}({\ddot{\mathbf{x}}_{j}})^{\mu }=\frac{e_{j}}{c}\left( \sum _{k\neq j}F_{k}|_{\textrm{ret}}\right) ^{\mu \upsilon }\left( {\dot{\mathbf{x}}_{j}}\right) _{\upsilon },\; \; j=1,\, 2,
\end{equation}
 where
\begin{equation}
\label{2.11}
F_{k}^{\mu \upsilon }=2c\int ^{\tau }_{-\infty }\left( {\dot{\mathbf{x}}}_{k}^{\upsilon }\partial _{\upsilon }-{\dot{\mathbf{x}}}^{\mu }_{k}\partial _{\mu }\right) \delta \left( \left( {\dot{\mathbf{x}}}_{j}(\tau )-{\dot{\mathbf{x}}}_{k}(\tau )\right) ^{2}\right) d\tau '.
\end{equation}

The features peculiar to this formulation can best be delineated by comparison
with Fokker's. The most outstanding difference is that Fokker's formulation
does not exploit Eq. (\ref{2.3}) and therefore employs a separate independent
parameter for each particle, which leads to a number of problems, including
synchronization of these parameters.\cite{Tru97} Fokker's Lagrangian is not
simply the sum of individual patched together in an \emph{ad hoc} manner; he
argued that a truly fundamental formulation should proceed from the variation
of a \emph{single} system Lagrangian to a set of coupled equations of motion.
the Lagrangian \( \cal {L}_{F} \),

\begin{eqnarray}
{\cal {L}_{F}}=\sum _{j}^{N}L_{j}=\sum ^{N}_{j}m_{j}\left( \mathbf{v}_{j}\cdot \mathbf{v}_{j}\right) ^{1/2} & - & \nonumber \\
2\sum ^{2}_{k\neq j}e_{j}e_{k}\int ^{+\infty }_{-\infty }\mathbf{v}_{j}(\tau _{j})\cdot \mathbf{v}_{k}(\tau _{k})\delta \left( \left( \mathbf{x}_{j}(\tau _{j})-\mathbf{x}_{k}(\tau _{k})\right) ^{2}\right) d\tau _{k}, &  & \label{3.1} 
\end{eqnarray}
 satisfies these criteria and leads, by means of the variation
\begin{equation}
\label{3.2}
\delta \int \sum ^{N}_{j}L_{j}d\tau _{j}=0,\; \; j=1,\, 2,\ldots ,\, N
\end{equation}
 to the equations of motion 
\begin{eqnarray}
m_{j}({\ddot{\mathbf{x}}}_{j}(\tau _{j}))^{\mu } & = & \frac{e_{j}}{2c}\sum ^{N}_{k\neq j}\left( F_{k}|_{\textrm{ret}}+F_{k}|_{\textrm{adv}}\right) ^{\mu \upsilon }\left( {\dot{\mathbf{x}}_{a}}(\tau _{j})\right) _{\upsilon },\nonumber \\
j & = & 1,\, 2,\, \ldots N\label{3.3} 
\end{eqnarray}
 These equations, however, cannot be integrated by a local procedure as is obvious
is one imagines attempting a machine integration of the \( j \)-th equation
at a given value of \( \tau _{j} \). Such an integration; i.e., a calculation
of the an incremental extention of the world line for an incremental increase
in \( \tau _{j} \), requires knowledge of the \( k \)-th world line on the
forward light cone of the \( j \)-th particle, which, in order to be computed,
requires knowledge of the \( i \)-th world line on the forward light cone of
the \( j \)-th particle, but this portion of this orbit is yet to be computed,
etc., \emph{ad infinitum.} In effect, the solution is needed as initial data
in order to compute the solution in this way.

Of course, advanced interaction could be precluded by changing the upper limit
of integration in Eq. (\ref{3.1}) to \( \tau _{ij} \), where \( \tau _{ij} \)
is that value of \( \tau _{j} \) which includes only the retarded potential
from the \( j \)-th particle; however, as \( \tau _{ij} \) would then also
appear in Eq. (\ref{3.1}), it could be written as the sum of individual Lagrangians
and therefore would not qualify as a system Lagrangian.

Schemes can be imagined which circumvent this problem by some sort of global
approach; i.e., by seeking the whole solution at once. For example, perhaps
the solution could be found as the limit of a technique each successive step
of which gave a closer approximation to the entire world line. At present, however,
such techniques appear to have not been developed---Eq. (\ref{3.3}) are in
general numerically and analytically unsolvable.

Eq. (\ref{2.9}), on the other hand, can always be integrated by machine because
the information needed to compute each incremental increase of any world line
has already been computed. Also by imagining a machine calculation, it is clear
that if each particle's world line between the past and the future with respect
to the same but otherwise arbitrary light cone is given as initial data, then
the system of world lines can be extended by calculations indefinitely into
the future or the past. (Note, however, that retrodicting is not simply equivalent
to reversing \( \tau  \) because the active and passive ends of the interaction
do not thereby also exchange roles. In other words, this formulation with differential-delay
equations of motion but advanced interaction has an intrinsic `time arrow.')
Although this type of initial data is greater that the customary Cauchy data
\( \{\mathbf{x}(\tau _{a}),\: {\dot{\mathbf{x}}}(\tau _{a})\} \), it is a general
characteristic of differential-delay equations that Cauchy data are insufficient
to determine a particular solution as enough initial data must be given to span
the delay.\cite{Bel63}

\section{Conclusions}

The main results of this work are twofold: 1.) strong doubt is cast on the validity
of the notion that at a fundamental level, Nature is nonlocal; and 2.) it is
shown that invariant proper time has the logical integrity required in order
to be the independent parameter for special relativistic mechanics. While the
first conclusion is without empirical contest, the second must still be reconciled
with several experiments whose current interpretation seems to be in conflict. 

The human psyche being what it is, it is in exactly those areas where certain
knowledge is the least likely, that compensation perversely induces the strongest
convictions. While matters of ``sex, politics and religion'' deliver the least
contestable examples of this effect, fundamental physics, because of its ``deep''
reputation and cultural affiliation with virtually transcendental wisdom, runs
a close second. Further, it seems to this writer that `non-locality' and `asymmetric
aging' tempt many who would like to find a confirmation of mystical, religious
or just exotic-futuristic beliefs, to see exactly that in logic-defying phenomena.
Real, honest science, however, demands uncompromised logical consistency. Likewise,
real, honest philosophy does not revel in antilogy as a portal on the preternatural,
rather the opposite (see various contributions in, e.g., \cite{Weg99}, wherein
time contortions are found wanting). 

The arguments supporting our conclusions are strictly points of logic, rather
than physical analogies or intuitive \emph{Ans\"atze}. Internal logical consistency
is even more demanding a standard than empirical verification whenever interpretation
intervenes. Internal consistency is vital for developing meaningful theory;
it is well known that if an axiom set contains an inconsistent element, it is
possible to prove as correct any statement whatsoever. In fact the test for
consistency consists essentially of finding a statement that can not be proven
true. To many it seems doubtful that modern physics could pass such a test;
thus, the continued development of physics theories best proceeds only on a
basis purged of all antinomy among its basic definitions and hypotheses. This
is our aim.

\end{document}